# Visible-blind ZnMgO Colloidal Quantum Dot Downconverters expand Silicon CMOS Sensors Spectral Coverage into Ultraviolet and enable UV Band Discrimination

Avijit Saha,[a] Gaurav Kumar,[a] Santanu Pradhan,[a] Gauttam Dash,[b] Ranjani Viswanatha[b] and Gerasimos Konstantatos*,[a,c]

[a]ICFO-Institut de Ciencies Fotoniques, The Barcelona Institute of Science and Technology, 08860 Castelldefels (Barcelona), Spain.

[b]New Chemistry Unit, Jawaharlal Nehru Centre for Advanced Scientific Research, Jakkur, Bangalore - 560064 India

[c]ICREA-Institució Catalana de Recerca i Estudis Avancats, Passeig Lluís Companys 23, 08010 Barcelona, Spain

**Abstract.** Selective spectral detection of ultraviolet (UV) radiation is highly important across numerous fields from health and safety to industrial and environmental monitoring applications. Herein, we report a non-toxic, visible-blind, inorganic quantum dot (QD)-based sensing scheme that expands the spectral coverage of Silicon CMOS sensors into the UV, enabling efficient UV detection without affecting the sensor performance in the visible and UV-band discrimination. The reported scheme employs zinc magnesium oxide (ZnMgO) QDs with compositionally tunable absorption across UV and high photoluminescence quantum yield (PLQY) in the visible. The efficient luminescence and large stokes shift of these QDs have been exploited herein to act as an efficient downconverting material that enhances the UV sensitivity of Si-photodetector (Si-PD). A Si-PD integrated with the QDs results in a nine-fold improvement in photoresponsivity from 0.83 mA/W to 7.5 mA/W at 260 nm. Leveraging the tunability of these QDs we further report on a simple UV band identification scheme, using two distinct band gap ZnMgO QDs stacked in a tandem architecture whose spectral emission color depends on the UV-band excitation light. The downconverting stack enables facile discrimination of UV light using a standard CMOS image sensor (camera) or by the naked eye and avoids the use of complex optics.

Future consumer electronics are foreseen to transition from imaging to perception offering information that lies beyond simple visualization. This requires sensing and imaging technologies with expanded spectral coverage and multispectral image fusion. Visible, near-infrared, and recently short-wave infrared have been identified as key spectral windows for which complementary metal-oxide-semiconductor (CMOS) -compatible low-cost optical sensors have been intensively investigated in recent years.[1] [2] Besides infrared, ultraviolet (UV) light is of equal, if not greater, importance due to the positive as well as negative consequences the exposure to it may have. UV light represents photons with energies higher than visible and depending on those there are various categorizations across the UV range. Within this range UV radiation can be classified as UVA (320-400 nm), UVB (285-320 nm), and UVC (10-285 nm), while the former comprises further two subgroups namely, short wave UVA or UVA2 (320-340 nm) and longwave UVA or UVA1 (340-400 nm). On account of very high energy, UV radiation can be very dangerous, causing serious health risks in humans[3, 4] and other living organisms.[5] Particularly, UVC and UVB bands can cause immediate skin damage leading to skin cancer,[4] cataracts,[6] melanoma,[3], etc. While UVA1 is mostly safe for human beings, prolonged exposure to UVA2 can cause sunburn and indirect damage to microbes' DNA.[7] Besides the downsides, UV also offers many beneficial effects e.g. vitamin D synthesis in the human body and mental wellness,[8] disinfection,[9] water purification, space communication,[10], etc. The wide energy range expansion of UV, leading to different penetration depths of various parts of the radiation cause remarkably diverse effects on living bodies. It becomes evident that awareness over the intensity as well as the photon energy of UV radiation is of paramount importance not only for consumers but also for environmental sensing and monitoring, process control inspection that makes use of UV light, etc. For most applications, UV sensing requires simultaneous visible imaging to fuse images from the different spectral windows, yet visible CMOS cameras have optical filters integrated for colored imaging that prevents them from being used in UV sensing. Additionally, silicon sensors optimized for UV detection, do so at the expense of their performance in the visible/near-infrared. For UV-only sensing applications, typically III-V nitrides are employed as visible-blind detectors that are not monolithic to silicon CMOS.[11] Thus, a solution that integrates seamlessly UV sensing with currently available image sensors remains a challenge.

Previously reported UV radiation sensors are mostly based on electrical responses of different semiconducting materials or devices involving mechanically hard components, which makes them intricate and incompatible for widespread direct applications.[12] [13] Visual sensing strategies for UV sensing have focused on UV absorbing materials relying on fluorescence/phosphorescence[14] or photochromatic materials[15] [16] typically used for UV-index measurements by visual means or materials that have parasitic absorption in the visible, precluding their facile integration atop a visible image sensor.[17] [18] [19] Si-PD is most widely used for broadband photodetection from VIS to NIR due to high reliability, excellent responsivity, and low-cost fabrication process.[20] [21] However, the responsivity of the Si-PD in the UV region, specifically in the UVC and UVB bands is very limited due to shallow penetration depth (small wavelength), high reflection coefficient, and hot electron loss.[22] Enhancement of UV responsivity to obtain a UV enhanced Si-PD is quite challenging, motivating research efforts.[23-26] Solution-processed nanocrystals (in particular Pb-perovskites) with high PLQY have been integrated on silicon detectors as downconverting films to enhance their sensitivity in the UV yet at the expense of performance loss in the visible from absorption losses of the downconverter.[26] [27]

We instead took the view that in order to integrate a UV sensitive material atop a CMOS visible camera enabling Si-CMOS image sensors as UV and VIS sensitive, we would need a material that has a tunable absorption across the UV range of interest without any absorption in the visible, a very high PLQY with large Stokes shift to emit visible light upon UV radiation, and it is solution-processed, thus low-cost and CMOS compatible, and last but not the least be RoHS compliant suited to consumer electronics. To fulfill these requirements, we turned to the use of colloidal quantum dot technology based on high bandgap semiconductors. We employed Mg-doped ZnO CQDs with tunable emission across the UV range by changing the doping stoichiometry of the CQDs to achieve a tunable UV absorber. The emission in such doped systems stems from emissive midgap trap-states.[28] [29] [30] In this work, we developed a material comprising ZnO CQDs with controlled doping to achieve UV-absorption and VIS-emission tuneable sensitizers with high PLQY and very large Stokes shift. We then integrated those materials into thin films atop a commercially available Si-detector to achieve a 9-fold enhancement in UV sensitivity without altering the detector's performance in the VIS-NIR. Furthermore, capitalizing on the tunability of the material we then made thin film stacks on glass that operate as UV-band discriminator with a visible color emission depending on the range of excitation in the UV, enabling us naked eye sensing of UVA1, UVA2, and UVB/UVC bands. The excellent optical properties of these QDs, namely their high and tunable UV absorption, along with their high emission and transparency in the visible, allowed us to demonstrate the performance of our UV sensor even in weak light conditions.

ZnMgO QDs were synthesized as described in methods in which $Mg^{2+}$ content stoichiometrically varied from (0-25)%. Depending on the amount of Mg-salt taken for different ZnMgO QDs, henceforth depicted as ZnMgO-n, where n is the stoichiometric percentage of Mg-salt. The actual percentage of $Mg^{2+}$ in ZnMgO was determined from inductively coupled plasma optical emission spectrometry (ICP-OES) depicted in SI Table T1. It is evident that while the stoichiometric amount was varied from (3-25)% the actual $Mg^{2+}$ incorporation into the ZnO matrix is from (1.9-12.4)%, which is consistent with the optical properties. The optical absorption and photoluminescence spectra of various ZnMgO QDs dissolved in ethanol are depicted in Figure 1a. Both UV-VIS absorption and PL-spectra show bandgap tunability offered through $Mg^{2+}$ incorporation in ZnO resulting in a blue shift. It is evident from Figure 1a, that with the incorporation of $Mg^{2+}$ ion the bandgap of ZnMgO increases from 3.70 eV to 4.22 eV (SI Table T1) resulting in a change of emission color from yellow to blue as shown in Figure 1c. The variation of the band positions (conduction and valance band) with $Mg^{2+}$ insertion were determined from UPS spectroscopy (SI Figure S1) and the energy level diagram has been shown in Figure 1d. PL-emissions from the QDs were quantified by measuring absolute PLQY using an integrating sphere. Undoped ZnO shows both band edge emission and emission due to intrinsic defect states or oxygen vacancy as shown in Figure 1a, however, the absolute PLQY is low (42%) as observed earlier.[31] Interestingly, with $Mg^{2+}$ incorporation, the band edge emission suppressed completely while we observe a remarkable increase in the mid-gap state emission leading to a maximum PLQY around 95% as illustrated in Figure 1b. The origin of such intense emissions was studied using photoluminescence excitation (PLE) scan. The PL excitation scans for different QDs corresponding to their emission peaks are shown in Figure S2. The PL excitation spectra look very similar to their corresponding absorption spectra suggests that while emission mostly occurs through the mid-gap states, light absorption is dominated by the host semiconductor bandgap. In order to understand the effect of $Mg^{2+}$ incorporation on the exciton decay dynamics and the recombination pathways inside the QDs, we studied time-resolved PL (TRPL) spectroscopy (Figure 1e). It is evident from the figure that the average lifetime (few µS) measured at the intense PL peak monotonically decreases with the $Mg^{2+}$ incorporation. In such cases, it is well known that the impurity atoms introduce intra-gap state/states in between the conduction and valance band.[29, 32] Similarly, $Mg^{2+}$ insertion into the ZnO matrix creates well defined single atomic-like states due to hybridization with the host ZnO. Radiative emission through these $Mg^{2+}$ states is more efficient resulting in improved PLQY. Moreover, we observed that the excitonic decay is comparatively faster through the mid-gap states of ZnMgO QDs compared to the oxygen vacancy related trap states of ZnO. This suggests that $Mg^{2+}$ insertion reduces the defect related trap states, which leads to faster radiative lifetime with the increase of $Mg^{2+}$ contents. Transmission electron microscopy (TEM) images of different representative ZnMgO-n (n = 3, 10, 15, 25) QDs are shown in Figure S3(a-d), including the selected area electron diffraction (SAED) pattern, whereas Figure S3(e-h) shows the high-resolution TEM (HRTEM) images highlighting the crystallinity and lattice planes. It is noteworthy that while the excitonic Bohr radius of ZnO is approximately 2.34 nm, the particle sizes of different ZnMgO-n QDs remain almost similar about 3.5 ± 0.5 nm which suggests that the variation of the bandgap is primarily because of the formation of an alloy over quantum confinement effect. It is worth noting here that further incorporation of $Mg^{2+}$ into ZnO using higher amounts of Mg-salt to further blue shift their absorption onset, following this technique, was challenging. More than 25% Mg taken stoichiometrically (30-50%) before reaction results in a completely different material (ZnMgO-30) as shown in SI Figure S4, which does not demonstrate high PL.

To ascertain the structural characteristics of the QDs, X-ray diffraction (XRD) and Raman spectroscopy were performed. The XRD pattern (Figure S5a) shows that all ZnMgO-n QDs presents diffraction peaks corresponding to the wurtzite phase of bulk ZnO without any trace of impurity peaks, which confirms the phase purity of these QDs. We did not observe any significant shift of any diffraction peak due to $Mg^{2+}$ ion incorporation in ZnO, since both $Mg^{2+}$ (72 pm) and $Zn^{2+}$ (74 pm) have almost the same ionic radii. Furthermore, Raman spectra of all different ZnMgO-n QDs (Figure S5b) show that the peaks are in good agreement with the vibrational modes of bulk ZnO (RRUFFID R050419). The broadening of the Raman peak for the QDs that has been observed in our experimental data is due to the phonon confinement inside the QDs.[33] [34] Raman spectra do not present any other secondary peaks of impurities such a MgO which indicates the pure phase formation of ZnMgO QDs.

The property of strong absorption in UV and efficient emission in the VIS range makes these CQDs suitable for applications where the high VIS sensitivity of a material can be utilized to enhance its low sensitivity for shorter wavelengths. To demonstrate this, we have integrated these QDs to enhance the UV sensitivity of a commercial Si-PD by exploiting its downshifting photoluminescence property. We have spin-coated ZnMgO-5 QDs (120 nm) on top of a Si-PD (described in Expt. Section) as shown in the schematic Figure 2a. Figure 2a illustrates the working principle of the integrated Si-PD in which the active QD film acts as the efficient down-converter of light from a poor sensitive region to a higher sensitive region of the Si-PD. As shown, the deposited QD film on Si-PD glows with an intense visible bright emission under a UV lamp (excitation-280 nm), which subsequently triggers the visible sensitivity of the Si-PD to generate a much higher photocurrent signal in response to UV illumination as compared to what it would be for directly falling UV illumination, thus enhancing its sensitivity. Figure 2b shows the optical property of the QD film that exhibits strong absorption in the UV region and emission in the VIS region (PLQY-90%). The PL excitation spectrum collected at the emission peak wavelength looks very similar to the absorption spectra, suggesting the efficient down-conversion from all over the UV range to the VIS. We observe that the PLE and emission spectra of QDs do not change significantly from being in solution phase to film form as shown in SI Figure S6. Figure 2c shows the transmission spectra of the film. It is interesting to note that the QD film is highly transparent with an average transmittance of 95% in the 360-1100 nm wavelength range. This excellent downconversion efficiency with stable photoluminescence and high transparency in visible makes ZnMgO QDs a potential candidate for UV enhanced in Si-PD.

To demonstrate the concept, in Figure 2d, we have further compared the responsivity of the Si-PD before and after ZnMgO-5 deposition on it. The responsivity of a photodetector is a key figure of merit (a function of wavelength, λ), defined by the ratio of photocurrent ($I_{ph}$) and incident photon power ($P_{in}$) as follows $R(\lambda) = I_{ph}/I_{in}$. Figure 2d and Figure S7 show the comparison of the responsivity of bare and ZnMgO deposited on Si-PD over a broad energy range from 240 -1100 nm. The responsivity of the original photodiode is poor under UV-light in the spectral range from 240-310 nm, however, with the introduction of the ZnMgO QDs layer, the responsivity enhanced significantly in this region. To achieve the highest enhancement in the responsivity out of this scheme, we optimized the layer thickness of the deposited QD film. We have measured the thickness (measured using a Surface profilometer and crossectional SEM, SI Figure S8) from 60 nm (1 layer ) to 180 nm (3 layers) and observed that a thickness of about 120 nm (corresponding to 2 layers) results in the maximum enhancement in the responsivity in the UV range as shown in the Figure S7. Although the PLQY of the QD films with different thicknesses (60 nm -240 nm) are very similar (SI Table T2), the responsivity value decreases both below and above 120 nm film thickness. While the thinner layer (60 nm) lacks in the absorption of the UV light, the thicker layer (180 nm) suffers more loss of emitted visible light through the sides of the QD film resulting in lower responsivity in both cases, as compared to the optimized thickness i.e. 120 nm (2 layers). This loss of emitted light in the case of the thicker film could be attributed to waveguiding effects in the QD slab and the increase in internal reflection resulting in less efficient light out-coupling. Henceforth, we discuss the results obtained from the 2 layered Si-PD. It is apparent from Figure 2d that ZnMgO integration in Si-PD, shows about a 9-fold enhancement in the responsivity (i.e. from 0.83 mA/W to 7.5 mA/W at 260 nm) than that of the bare Si-PD, while not deteriorating the performance of the PD in the visible region. Figure 2e illustrates the relative increase in the responsivity of the QD-integrated PD compared to the bare one. We observed a record increase of about 800% at 260 nm indicating the significance of ZnMgO QDs as efficient down-converting materials for Si-PD. The absolute responsivity values reached herein are determined by the initial responsivity of the Si detectors at 260 nm that were used in this study. Thus, higher values are within reach by employing silicon detectors with higher responsivity in the UV. To the best of our knowledge, this relative enhancement (9-fold) in the responsivity in our device is the highest ever reported compared to similar recent studies.[24, 25] It is interesting to note that the enhancement curve (Figure 2e) looks very similar to the absorption spectrum of the ZnMgO film, which confirms that the enhancement is directly correlated to the sample's absorption and emission. However, the responsivity decreases slightly around 330-360 nm due to the inefficient light absorption by the QD layer together with the scattering effect. Furthermore, for wavelengths above 360 nm to 1100 nm, the QD film is almost transparent and the scattering reduced (due to inverse $\lambda^4$ dependence) results in very similar responsivity to bare Si-PD. Thus, this scheme allows us to introduce a UV enhanced Si-PD without affecting its responsivity in the VIS and NIR region.

However, there are some difficulties in applying luminescent materials as down-shifting materials in this process.[24-26] Although the reported QD integrated Si-PDs shows a significant enhancement of responsivity in the UV region, the light incident on the materials emits in all 360° direction uniformly, which results in more than 50% loss of the emitted light that radiates away from the detector, and thus undermines the complete potential. Moreover, complete transparency in the sensitive region of the photodetector is essential, otherwise, absorption in that region might reduce the resulting sensitivity of the device compared to the bared one. Therefore, in addition to the high absorption coefficient in the UV region, the material should also be transparent above the targeted UV region with a large stokes shift and high PLQY. After the recent advancement of high PLQY perovskite QDs, many groups have shown UV enhancement in Si-PD using perovskite QDs.[24] [26] The ZnMgO QDs employed here have strong absorption in the UV range and high transparency in the VIS range (Figure 2b, c) along with strong emission in VIS that occurs through the mid-gap states, resulting in a strong stokes shift. This leads to an emission overlap with the high-efficiency region of Si-PD giving rise to a record enhancement of responsivity in the UV region.

Leveraging the spectral tunability of the CQDs we sought to develop a simple yet effective approach to enable UV-band discrimination. This is of paramount importance in cases where information on the type of UV radiation exposure is crucial for process control monitoring, safety, etc. Figure 3a represents a schematic of a passive UV-band discriminator (specific to UVA and UVB/UVC). The mechanism for the UV sensor involves selective absorption and transmission of specific UV bands across the layer stack that results in specific colored luminescence. The sensor combines two different QDs layers with distinct bandgaps in a tandem fashion with optimized thicknesses. We observe (SI

Table T1) that ZnMgO-3 QDs with low $Mg^{2+}$ content have a comparatively smaller bandgap (3.7 eV) absorbing light from across the UV bands while ZnMgO-25 QDs with more $Mg^{2+}$ content, have a significantly larger bandgap (4.2 eV). Therefore, the ZnMgO-25 layer absorbs only higher energy bands i.e. UVB and UVC, and is nearly transparent to UVA light. This inspired us to design a sensor that can selectively sense the different UV bands owing to the tunability of the absorption of ZnMgO QDs. To show this, here, we have deposited ZnMgO-3 as a bottom layer (80 nm) on a glass substrate and ZnMgO-25 as the top layer (280 nm) and optimized the thickness of each layer to achieve the optimized performance as illustrated in Figure 3a. A comparatively thicker top layer of ZnMgO-25 was used so that it can absorb most of the UVB light under exposure. Since ZnMgO-25 is almost transparent to UVA, it acts like a long-pass filter as shown in Figures 3a and 3b letting UVA pass through it with minimal attenuation and mostly absorbed by the bottom ZnMgO-3 QD film. This scheme thus provides the required selectivity in sensing the specific UV bands. Furthermore, as each type of QDs has its signature emission which is visibly distinct by the virtue of the large stokes shift and uniform PL shift, the proposed tandem structure results in yellow emission from ZnMgO-3 and blue emission from ZnMgO-25 layers. In addition to that, high PLQY offers a strong and unique visual response under different UV exposure in this device and thus enables naked-eye detection. We measured the excitation-dependent PL-Spectra from the device as depicted in Figure 3c which shows that the emission peak position remains constant at 486 nm with the excitation wavelength below 320 nm while excitation from 340 nm results in an emission peak at 555 nm (Figure 3c, d). However, due to the presence of the absorption tail of the ZnMgO-25 QDs, both QD layers absorb in the range from 320-340 nm resulting in a mixed emission peak between 486-555 nm. It is evident from the figure that with the increasing excitation wavelength from 320 nm to 340 nm the emission peak also shifts towards the longer wavelength due to dominant contribution from the bottom ZnMgO-3 layer. The corresponding digital photographs of the sensor device taken under different excitation wavelengths demonstrate a naked eye distinguishable colorimetric response, producing blue color for UVB and yellow color for UVA1 exposure. It is interesting to note that for 320-340 nm wavelength range (Figure 3b), the partial absorptions of both the QDs layer leads to mixing of both blue and yellow emission resulting in green emission (photograph in Fig 3c, d) in this small window, which allows us to detect the shortwave UVA2.

Thus, the sensor provides unique color responses for different ranges of UV exposure as shown in the digital photographs in Figure 3c enabling visual detection of various UV bands. Furthermore, for the implementation of this sensor in real-world applications to alert the user even under low UV light exposure, it is imperative to have an ultrasensitive response from the sensor. In this device, the response depends on the photon down-conversion efficiency or the PLQY of the QD films that have been used here. Hence, power-dependent response demonstrating (Figure S9) the sensor's ability to differentiate UV bands under low light exposure has been investigated. Here, we have studied the sensor's response only under varying UVA2 and UVA1 light due to the inaccessibility of the appropriate UV filter. However, it is evident from the digital photographs (shown in Figure S9) that the sensor continues to show naked-eye distinguishable response even under very low power (10-30 $\mu W/cm^2$) of UV irradiation. Thus, based on the use of selective absorption and transmission of UV radiation by different UV absorbing QDs and their distinct emission properties, a simple yet novel, spectrally selective, highly sensitive visual UV band sensor has been demonstrated. Moreover, it is also important for the QDs to have significant and stable PLQY on film, should retain it at high temperature and for a prolonged time at room atmosphere. Interestingly, here the high PLQY of the ZnMgO QDs does not decrease significantly from the solution phase (95%) to film (90%). This is due to two key factors (i) large stokes shift which suppresses self-absorption in QDs film and Forster energy transfer, (ii) usually ligand exchange during film growth results in defects due to improper surface passivation, leading to low PLQY in films, however, the film growth without ligand exchange here allowed us to retain the same surface environment around the QDs in both solution and film. Furthermore, we studied the temperature-dependent (SI Figure S10) and temporal (SI Figure S11) stability of absolute PLQY of undoped ZnO and ZnMgO QDs films annealed in air at various temperatures and observed that the ZnMgO QDs films have high optical stability. This robust PL-efficiency on the film suggests the in-principle feasibility of employing ZnMgO for realistic UV sensing application.

In summary, this work reports the synthesis of environmentally-friendly non-toxic ZnMgO QDs with composition tunable bandgap and excellent optical properties of tunable, near unity QY (95%), high transparency in VIS region, and large stokes shift. The excellent optical properties of ZnMgO QDs have been exploited for a suitable downconverter application. Integrating these QDs on top of the Si-PD, we have successfully fabricated a low-cost device that exhibits UV-enhanced broadband photodetection without affecting the VIS and NIR responsivity. The use of an optimized QDs film resulted in a remarkable relative enhancement of 800% in the responsivity of the Si-PD in the UV region of 240-330 nm. Furthermore, utilizing these QDs, for the first time, zero-power, visual, and spectral-sensitive UV sensor based on environmentally friendly semiconductor material, has also been demonstrated. The sensor employs a simple concept to use two layers of QDs having distinguished optical properties that establish a unique ability to spectrally differentiate UVB, UVA1, and UVA2 radiations. This work paves a new pathway for utilization of environmentally friendly QDs for selective UV radiation sensing as well as triggering UV enhanced photosensitivity of Si-PD.

**Experimental Section:**

**Materials:** Zinc acetate dihydrate (Zn $(Ac)_2$,$2H_2O$) 99%, magnesium acetate tetrahydrate (Mg $(Ac)_2$, $4H_2O$) 98% were purchased from Sigma Aldrich. Dimethyl sulfoxide (DMSO) 99.9%, potassium hydroxide (KOH) pellets were purchased from Scharlab. Anhydrous Ethanol absolute was purchased from VWR chemicals. All these chemicals were used without further purification.

**Synthesis:** $Zn_{1-x}Mg_xO$ QDs were synthesized after modification of the reported method in the literature.[35] In a typical synthesis required amount of Zn $(Ac)_2$, $2H_2O$, and Mg $(Ac)_2$,$4H_2O$ and 30 ml of DMSO was taken in a three-necked round-bottom flask equipped with a temperature controller probe and heating mantle. The total amount of zinc and magnesium salt was kept constant to 3 mmol. The reaction mixture was stirred for 15 min at room temperature. At the same time, in a separate vial, 5 mmol of KOH pellets were dissolved in 10 ml of anhydrous ethanol solution. The prepared KOH solution was then added dropwise to the reaction mixture inside a three-necked round bottom flask solution for a period of 2 min at a temperature of 30 °C with constant stirring. The reaction conditions were left unaltered and kept with continuous stirring in ambient conditions. After 25 min the temperature of the reaction mixture was raised to 70 °C. As soon as it reaches 70 °C the temperature was cooled down to room temperature by removing the heating mantle and using a water bath. The as-prepared nanocrystals were washed and purified twice using Acetone/Ethanol mixture and centrifuged to obtain nanocrystal precipitation and re-dissolved in Ethanol. Finally, the dispersion was centrifuged at 5000 rpm for 4 min and the supernatant was collected the larger particles were discarded. Undoped ZnO nanocrystals were synthesized following the same condition in absence of magnesium salt.

**Characterizations.** Absorption spectra of samples were recorded using Carry 5000 UV-Vis- NIR spectrophotometer while steady-state PL and PL-excitation spectra were obtained using Horiba PL1057 spectrophotometer. Absorption and transmission were performed by taking the sample in a quartz cuvette which is transparent up to 200 nm and measured using an integrating sphere. Absolute QY, both in solution and on film, was determined using an integrating sphere attached to the PL-spectrophotometer. TRPL spectra were obtained from QDs film using a microflash lamp (Excitation wavelength 270 nm) as the source on the FLSP920 spectrometer, Edinburgh instrument. TEM was carried out using JEOL 2100 microscope using a field emission gun (FEG) operating at an acceleration voltage of 200 kV. X-ray diffraction patterns for the nanocrystals were recorded on a Regaku Alpha1 diffractometer using Cu-Kα ($\lambda$=1.5406Å) radiation. Raman spectroscopy measurements were carried out on a Renishaw InVia spectrometer using a 532 nm laser at room temperature. An objective lens (50x) was used to focus the lasers on the samples. Samples for UPS were prepared on ITO substrates by spin coating (2500 rpm) ZnMgO-n QDs solution (40 mg/ml dispersed in Ethanol). UPS measurements were performed at the Institut Català de Nanociencia i Nanotecnologia (ICN2) on a SPECS PHOIBOS 150 electron spectrometer using monochromated HeI radiation (21.2 eV) with a biasing voltage of 10 eV. Elemental analysis of the samples was carried out using inductively coupled plasma optical emission spectroscopy (ICP-OES) Perkin Elmer, model Optima 3200RL in standard conditions.

**Device.** UV band sensing unit was prepared on a glass substrate. A 40 mg/ml ZnMgO-3% QDs was spin-coated on a glass substrate with a speed of 2500 rpm for 60 s which grows 80 nm thick film. After that 50 mg/ml ZnMgO-25% QDs were spin-coated three times with layer by layer processed to grow further 280 nm of these blue-emitting QDs. In the above process, the glass substrate was annealed at 80 °C in the air for 60 s after each layer formation. The thickness of the film was measured using a KLA Tencor Alpha-Step IQ surface profilometer and scanning electron microscopy (SEM). The sensor was illuminated with all different excitation wavelengths ranging from UVC to UVA, inside the Horiba PL spectrophotometer. Photos under different excitation were taken with a digital camera. To study the power dependence, we made an experimental set-up. The monochromatic UV light from the Xenon lamp was guided through an optical fiber and passed by a monetarized filter wheel equipped with neutral density (ND) filters having different optical densities. The sensor was placed after that facing the sample side towards the source. The power density of the incident light was controlled by changing the wheel filters and was measured using calibrated Newport 818-UV/DB Si-PD connected to a Newport 1918-C power meter.

Commercial Si-detector was bought from ON Semiconductor and the detectors were attached to a glass substrate using two-sided tape. 30 mg/ml ZnMgO-5% QDs solution dissolved in ethanol spin-coated on the Si-detector with a speed of 2500 rpm. Thickness was varied by repeating this process and drying the detector in the air for 5-10 min after each layer growth. For optical studies and thickness determination, we imitate the QDs film on a quartz substrate. We used a 450 W Xenon arc lamp as the light source which guided through a monochromator for wavelength tuning. The output light was guided through an optical fiber and focused on the Si-PD. The output power over all the measured wavelengths of the light coming out from the optical fiber was measured using a calibrated standard Si-PD and power meter as described before. The photocurrent from the bare and QD integrated Si-PD was measured using Keithley 2400 source measurement unit.

## ACKNOWLEDGMENT

The authors acknowledge financial support from the European Research Council (ERC) under the European Union's Horizon 2020 research and innovation programme (grant agreement no. 725165), the Spanish Ministry of Economy and Competitiveness (MINECO) and the 'Fondo Europeo de Desarrollo Regional' (FEDER) through grant TEC2017-88655-R. The authors also acknowledge financial support from Fundacio Privada Cellex, the program CERCA and from the Spanish Ministry of Economy and Competitiveness through the 'Severo Ochoa' Programme for Centres of Excellence in R&D (SEV-2015-0522). We thank Zhuoran Wang for helping in taking SEM images.

**Supporting information** Supporting Information is available from the Wiley Online Library or from the author.

**Figures:**

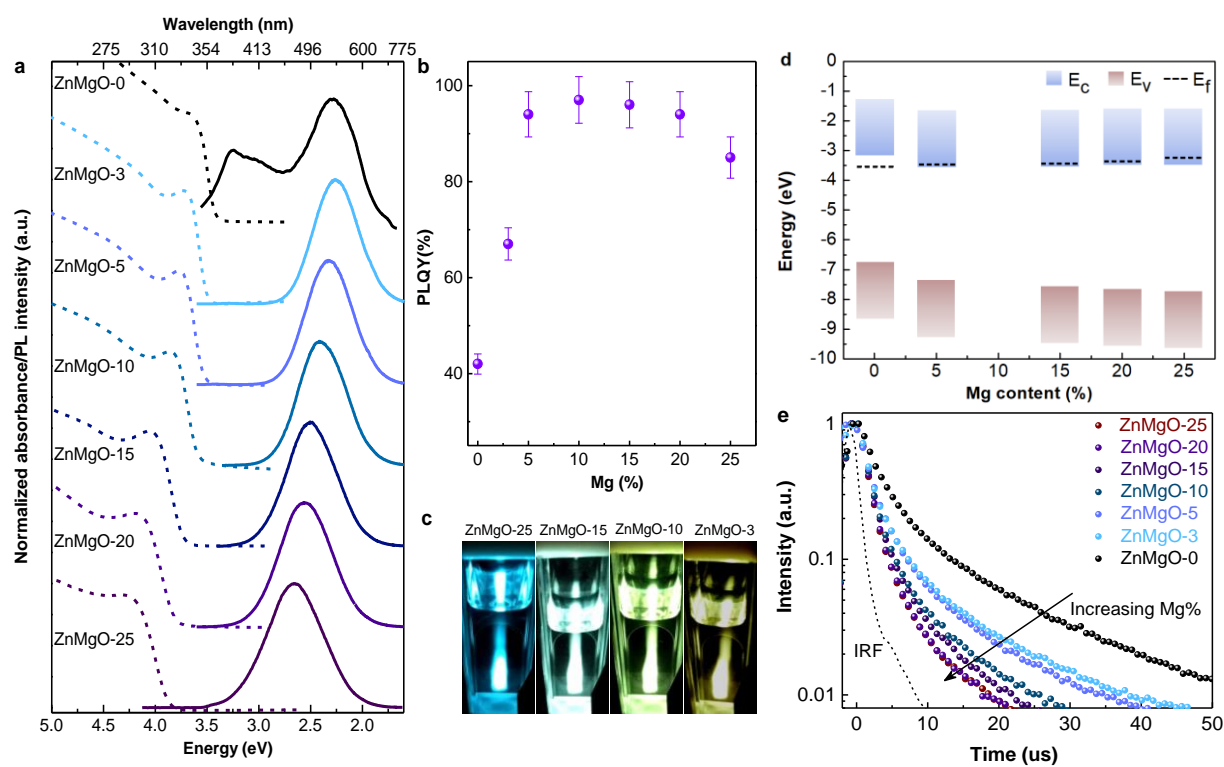

**Figure 1. Optical properties of ZnMgO-n QDs. a** Absorption (dashed line) and photoluminescence (solid line) spectra of ZnMgO-n QDs dispersed in ethanol with stoichiometrically increasing Mg content (n= 0, 3, 5, 10, 15, 20, 25) from top to bottom. **b** PLQY's of QDs in solution phase measured using the integrating sphere with an excitation wavelength 290 nm (for ZnMgO-25 and ZnMgO-25) and 300 nm for all other samples. **c** Photographs of the QDs in ethanol under weak steady-state UV light exposure (300 nm). **d** Energy band diagram for ZnMgO-n QDs with different Mg amounts. **e** TRPL spectra of various ZnMgO-n QDs collected at their corresponding emission peak position with an excitation wavelength of 270 nm. The black dotted line represents the instrument response function (IRF).

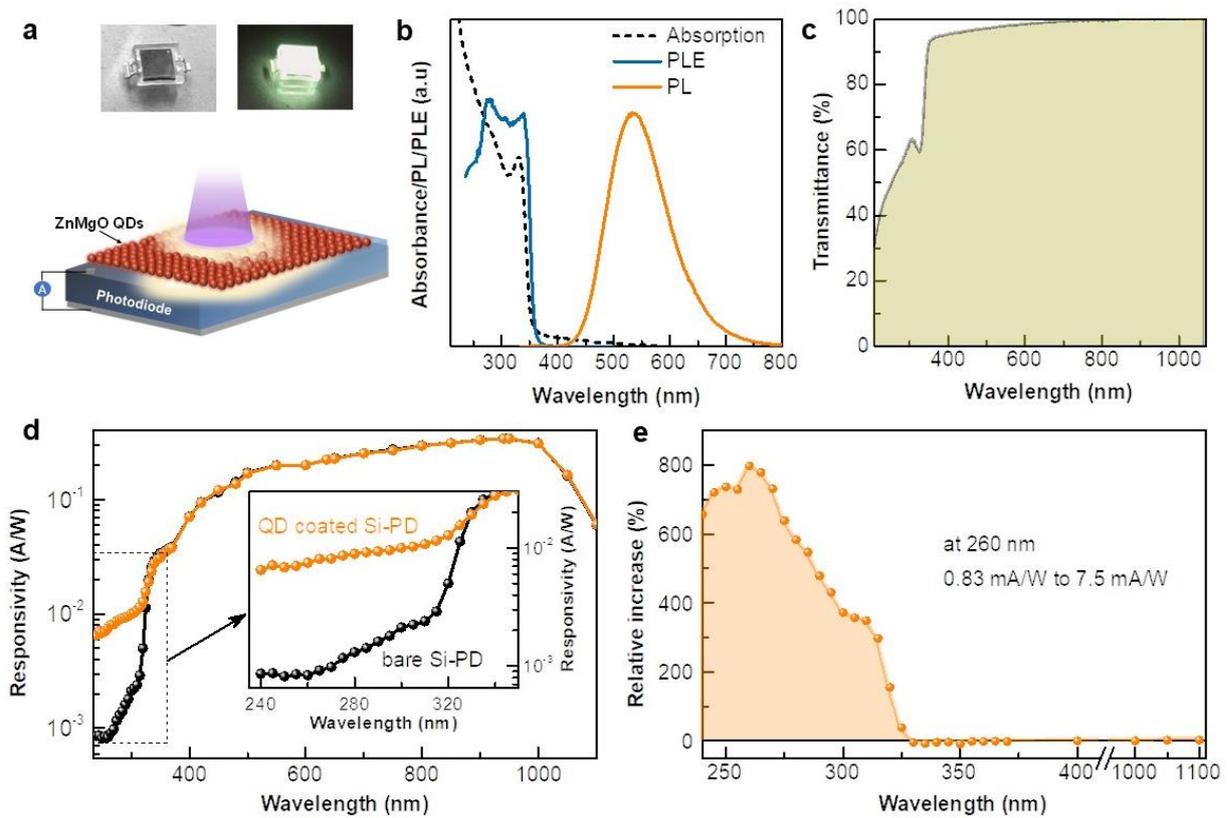

**Figure 2. Photophysical properties and performance of QD integrated (120 nm) Si-PD. a** Schematic illustration of the configuration and working principle of UV enhanced Si-PD, along with a digital photograph of QD integrated Si-PD under ambient light (left panel) and UV light (280 nm, right panel). **b** Optical absorption (black dotted line), PL excitation scan (blue line) at emission peak 540 nm, and PL emission (orange line, excited at 280 nm) spectra of QD film. **c** transmission spectra of QDs film taken using an integrating sphere. **d** Comparison of the responsivity spectra for the bare (black) and QD-thin film covered Si-PD (orange). The inset shows the enlarged view of the responsivity curve in the UV region. **e** Relative increase in the responsivity of the QDs integrated Si-PD.

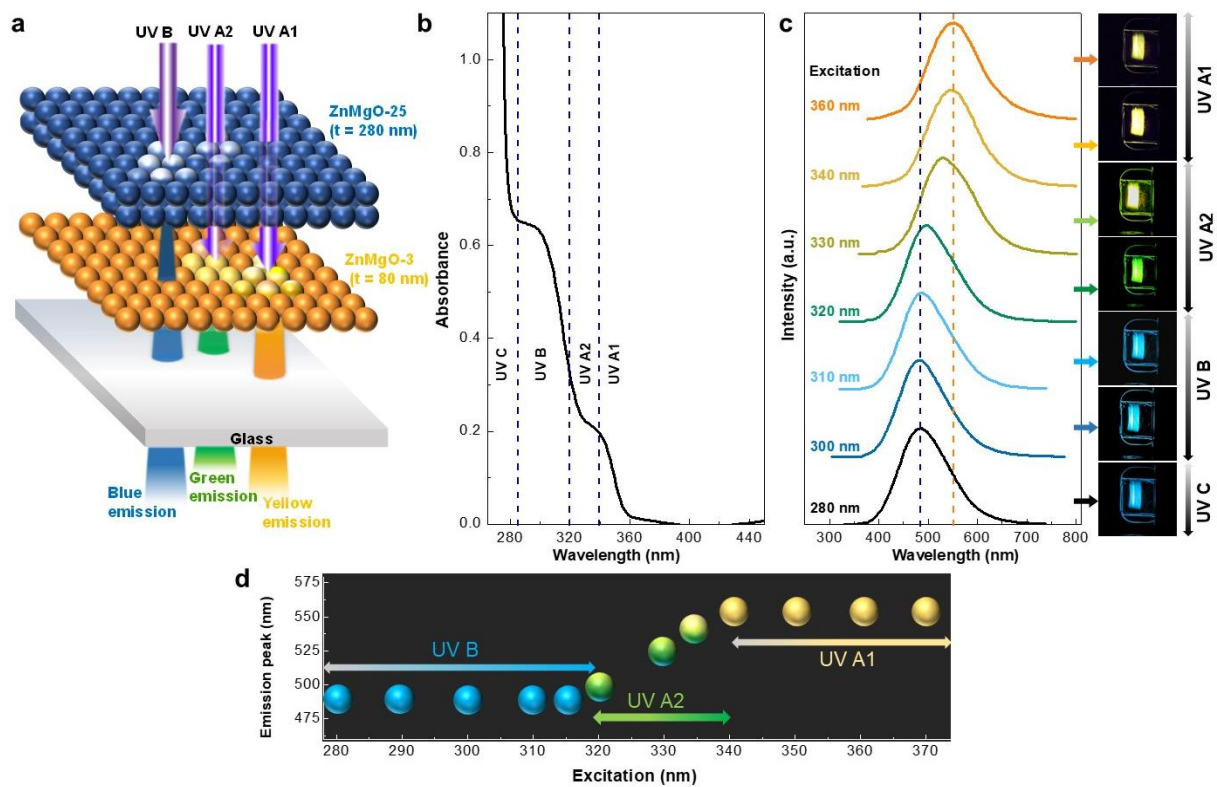

**Figure 3**. **Working principle of UV band sensor**, **a** Schematic illustration of zero-power UV band sensor and its principle of operation. **b** Absorption spectra of the sensing film. In order to take the absorption, we have imitated the device on a quartz substrate. **c** photoluminescence spectra of the sensor with different excitation and their corresponding photograph. For both photographs and PL measurement, the sample was faced on the excitation source side. **d** Schematic plot of excitation wavelength vs emission peak position. The color of the sphere represents the color of emission from sensing film where it shows blue, green, and yellow emission for UVB, UVA2, and UVA1 exposure respectively.